\documentclass[a4paper,fleqn,usenatbib]{mnras}

\usepackage[T1]{fontenc}
\usepackage{ae,aecompl}

\usepackage{graphicx}	
\usepackage{amsmath}	
\usepackage{amssymb}	
\usepackage{txfonts}
\usepackage{enumitem}
\usepackage{natbib}
\usepackage{xcolor}
\usepackage{multirow}
\usepackage{pdflscape}

\title[Atmospheric dispersion]{On-sky measurements of atmospheric dispersion: \\ II. Atmospheric models characterization}

\author[B. Wehbe et al.]{
B. Wehbe,$^{1,2}$\thanks{E-mail: bachar.wehbe@astro.up.pt}
A. Cabral,$^{3,4}$
L. Sbordone$^5$,
and G.\'{A}vila
\\
$^{1}$Instituto de Astrof\'{i}sica e Ci\^{e}ncias do Espa\c{c}o, Universidade do Porto, CAUP, Rua das Estrelas, 4150-762 Porto, Portugal\\
$^{2}$Departamento de F\'{i}sica e Astronomia, Faculdade de Ci\^{e}ncias, Universidade do Porto, Rua Campo Alegre, 4169-007 Porto, Portugal\\
$^{3}$Instituto de Astrof\'{i}sica e Ci\^{e}ncias do Espa\c{c}o, Universidade de Lisboa, Campus do Lumiar, Estrada do Pa\c{c}o do Lumiar 22, Edif. D, PT1649-038 Lisboa, Portugal\\
$^{4}$Departamento de F\'{i}sica, Faculdade de Ci\^{e}ncias, Universidade de Lisboa, Campo Grande 1749-016 Lisboa Portugal \\
$^{5}$European Southern Observatory, Alonso de C\'{o}rdova 3107, Vitacura, Regi\'{o}n Metropolitana, Chile
}

\date{Accepted XXX. Received YYY; in original form ZZZ}

\pubyear{2020}

\begin{document}
\label{firstpage}
\pagerange{\pageref{firstpage}--\pageref{lastpage}}
\maketitle

\begin{abstract}
Differential atmospheric dispersion is a wavelength-dependent effect introduced by Earth's atmosphere that affects astronomical observations performed using ground-based telescopes. It is important, when observing at a zenithal angle different from zero, to use an Atmospheric Dispersion Corrector (ADC) to compensate this atmospheric dispersion. The design of an ADC is based on atmospheric models that, to the best of our knowledge, were never tested against on-sky measurements. We present an extensive models analysis in the wavelength range of 315-665 nm. The method we used was previously described in the paper I of this series. It is based on the use of cross-dispersion spectrographs to determine the position of the centroid of the spatial profile at each wavelength of each spectral order. The accuracy of the method is 18 mas. At this level, we are able to compare and characterize the different atmospheric dispersion models of interest. For better future ADC designs, we recommend to avoid the Zemax model, and in particular in the blue range of the spectra, when expecting residuals at the level of few tens of milli-arcseconds.
\end{abstract}




\section{Introduction}
\begin{figure}
	\centering
	\includegraphics[width=\hsize]{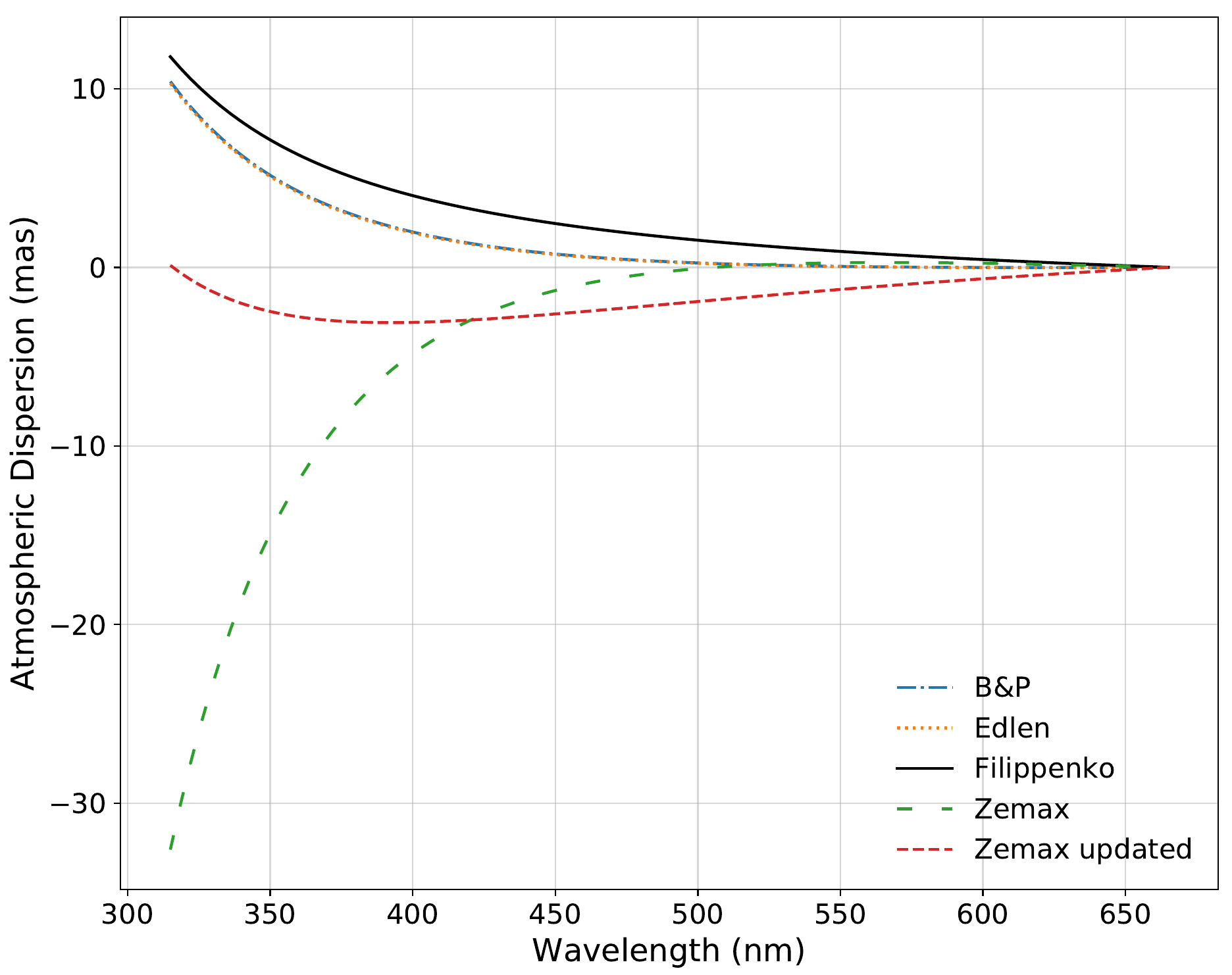}
	\caption{Differential atmospheric dispersion at Paranal for different models between 315 nm and 665 nm. The average between the models was used as reference.}
	\label{Fig:model_diff}
 \end{figure}

Differential atmospheric dispersion is a wavelength-dependent effect introduced by Earth's atmosphere that affects astronomical observations performed using ground-based telescopes. It is important to take this atmospheric dispersion into consideration, when observing at a zenithal angle different from zero. The common way to compensate this dispersion, is to implement an Atmospheric Dispersion Corrector (ADC) to compensate the on-sky dispersion. Most of ADCs are counter-rotating prisms used to correct the effects of atmospheric dispersion \citep{Cabral2012}. Each prism is a set of two prisms at least, made from different glasses. The choice of glasses is based on their refractive index, and their ability to reproduce the same amount of dispersion that needs to be corrected. The design of an ADC is based on atmospheric models used to estimate the on-sky dispersion to be corrected. During this phase, several pairs of glasses are chosen and tested. The final selection of glass is usually based on the minimum amount of residuals (on-sky dispersion - ADC dispersion) they can provide. This can be considered a good function of merit if the level of accuracy of the atmospheric dispersion models is lower than the level of residuals desired. High-precision astronomical instruments, like state-of-the-art spectrographs, are now equipped with ADCs that are supposed to deliver residuals at the level of few tens of milli-arcseconds (mas). Having residuals below 100 milli-arcseconds (mas), for objects that have a full width at half maximum (FWHM) of 1000 mas is quite challenging \citep{Wehbe2020a}. We notice that the difference between the most used atmospheric models, \citet{Filippenko1982}, hereafter Filippenko, and \citet{Zemax}, hereafter Zemax, can reach $\sim$ 50 mas, the limit of residuals needed (see Figure \ref{Fig:model_diff}). Designing ADCs to return residuals at the same level of the difference between the models will introduce errors that will affect flux losses and the final radial velocity (RV) computed as shown in \cite{Wehbe2020a}. Therefore, it is important to characterize different atmospheric models and quantify their accuracy in terms of reproducing the on-sky dispersion down to the level of few mas. \\
Several attempts to theoretically compare different atmospheric models can be found in the literature \citep[e.g.,][]{Spano2014}. To the best of our knowledge, these models have never been tested on-sky. In the first paper of this series \citep{Wehbe2020b}, we presented a method to measure on-sky the atmospheric dispersion. In Section \ref{sec:models}, we describe the models we are interested in. We also present a qualitative analysis of the different models. The method is summarized in Section \ref{sec:method_observations}, together with the different observations used in this work. The data analysis is presented in Section \ref{sec:data_analysis}, and the conclusions in Section \ref{sec:conclusions}.

\section{Models}
\label{sec:models}
The effects of atmospheric dispersion are well known and have been extensively discussed by several authors \citep[e.g.,][]{Filippenko1982, Skemer2009, Bechter2018, Blackman2019, Wehbe2020a}. Several attempts were done to measure the refractivity of dry air, $n$ , using an air refractometer \citep[for example][hereafter B\&P]{Bonsch1998} , and derive a formula to compute the refractivity as a function of wavelength. Using the air refractivity formula, we can derive the amount of atmospheric dispersion using the following:
\begin{align}
	\Delta R (\lambda) &= R(\lambda) - R(\lambda_{\rm ref}) \nonumber \\
	\Delta R (\lambda) &\approx 206265 \left[n(\lambda) - n(\lambda_{\rm ref}) \right] \times \tan Z,
	\label{eq:dispersion}
\end{align}
where $R(\lambda)$ is the refraction angle, $\lambda_{\rm ref}$ is the reference wavelength, and $Z$ is the zenithal angle of observation. Since observatories are usually located at an altitude higher than sea level, the atmospheric parameters, such as temperature, pressure and relative humidity (RH), are also taken into consideration while deriving the air refractivity formula. Different attempts to measure the refractivity using in-lab measurements, give different results. This difference is minimized when taking into account the variation of pressure, temperature and RH. Another important difference is the limits of the wavelength range where their formulaf are trustable. \\
In this section, we will briefly describe the different models of interest and compare them in a qualitative way.

\subsection{Edlen}
A dispersion formula for standard air (15$^{\circ}$ and 760 mbar) was derived by \cite{Edlen1953} based on measurements done by \cite{Barrell1939}. In 1966, an improved version of the dispersion formula was provided by \cite{Edlen1966}, hereafter Edlen, that took into account the carbon dioxide and water vapor. The main dispersion formula is given by equation \ref{eq:ed_s}, followed by the effects of carbon dioxide, temperature and pressure, and the water vapor, respectively.
\begin{equation}
    \label{eq:ed_s}
    (n-1)_{\rm s} \times 10^{8} = 8342.13 + \frac{2406030}{130-(1/\lambda)^2} + \frac{15997}{38.9-(1/\lambda)^2}
\end{equation} 

\begin{equation}
    \label{eq:ed_x}
    (n-1)_{\rm x} = [1+0.54(x-0.0003)](n-1)_{\rm s}
\end{equation}

\begin{equation}
    \label{eq:ed_tp}
    (n-1)_{\rm tp} = \frac{p(n-1)_{\rm s}}{720.775} \times \frac{1+p(0.817-0.0133t) \times 10^{-6}}{1+0.003661t}
\end{equation}

\begin{equation}
    \label{eq:ed_tpf}
    n_{\rm tpf} = n_{\rm tp} - f(5.722-0.0457(1/\lambda)^2) \times 10^{-8},
\end{equation}
where $x$ is the percentage of carbon dioxide, $f$ is the water vapor pressure, $t$ is the temperature, and $p$ is the atmospheric pressure. Combining equations \ref{eq:ed_s}, \ref{eq:ed_x}, \ref{eq:ed_tp}, and \ref{eq:ed_tpf} with equation \ref{eq:dispersion}, result with the atmospheric dispersion using Edlen's formula. \\


\subsection{Filippenko}
\cite{Filippenko1982} introduced a new set of formula to compute the air refractivity of standard air (15$^{\circ}$ and 760 mbar). He started using the 1953 Edlen's formula \citep{Edlen1953} and updated the dependence on pressure, temperature, and water vapor (Equations \ref{eq:f_f} to \ref{eq:f_tp}).

\begin{equation}
    \label{eq:f_f}
    n_{\rm f} = f \times \frac{0.0624-0.00068(1/\lambda)^2}{1+0.003661t}
\end{equation}

\begin{equation}
    \label{eq:f_s}
    (n-1)_{ \rm s} \times 10^{6} = 64.328 + \frac{29498.1}{146-(1/\lambda)^2} + \frac{255.4}{41-(1/\lambda)^2} - n_{\rm f}
\end{equation}

\begin{equation}
    \label{eq:f_tp}
    (n-1)_{\rm tp} = (n-1)_{\rm s} \times \frac{p[1+(1.049-0.0157t)p \times 10^{-6}]}{720.883(1+0.003661t)}
\end{equation}

\subsection{B\"{o}nsch \& Potulski}

\cite{Bonsch1998} updated the Edlen's formula to adapt the change of systems where the standard air is now considered at 20$^{\circ}$ and 750 mbar. They also built an air refractometer and were able to measure the refractivity. They refined their formula using the best fit of their results. In a similar way, we will list the main equations of B\&P to compute the atmospheric dispersion (Equations \ref{eq:bp_s} to \ref{eq:bp_tpf}).

\begin{equation}
    \label{eq:bp_s}
    (n-1)_{\rm s} \times 10^{8} = 8091.37 + \frac{2333983}{130-(1/\lambda)^2} + \frac{15518}{38.9-(1/\lambda)^2}
\end{equation}

\begin{equation}
    \label{eq:bp_x}
    (n-1)_{\rm x} = [1+0.5327(x-0.0004)](n-1)_{\rm s}
\end{equation}

\begin{equation}
    \label{eq:bp_tp}
    (n-1)_{\rm tp} = \frac{p(n-1)_s}{93214.6} \times \frac{1+p(0.5953-0.009876t) \times 10^{-8}}{1+0.003661t}
\end{equation}

\begin{equation}
    \label{eq:bp_tpf}
    n_{\rm tpf} = n_{\rm tp} - f(3.802-0.0384(1/\lambda)^2) \times 10^{-10},
\end{equation}

\subsection{Zemax}
The Zemax model of atmospheric dispersion is based on the work of \cite{Zemax}. The main refractivity formula is based on the one published by \cite{Barrell1939}. The dependence on temperature, pressure, and RH, is computed in a complex way to take into account all the variations in the troposphere and stratosphere layers of the atmosphere. Therefore, we will only mention the main refractivity formula here (Equation \ref{eq:z_s}). For more details, the reader may refer to \cite{Barrell1939}, and \cite{Zemax}.

\begin{equation}
    \label{eq:z_s}
    (n-1)_{\rm s} \times 10^{8} = 27263.24 + \frac{154.4}{\lambda ^2} + \frac{1.29}{\lambda ^4}
\end{equation}

\subsection{Zemax updated}
\cite{Spano2014} introduced an improved version of the Zemax atmospheric dispersion model. The Zemax model underestimates the atmospherisc dispersion in the UV and the in the blue part of the visible spectrum by 0.07 \%. Therefore, the author introduced a thin prism before the Zemax model to better describe the air refraction with a ``gradient 5" surface. The prism apex angle is scaled as a function of the observed zenithal angle Z (see Equation \ref{eq:zu_prism}). The reader may refer to \cite{Spano2014} for more details.

\begin{equation}
    \label{eq:zu_prism}
    \tan \delta = \frac{1}{G} \left[A \tan Z + B (\tan Z)^3 \right],
\end{equation}
where G is a scale factor used to define the prism index N, $\rm A ~ = ~ 1.00472$, and B = -0.00103.

\subsection{Qualitative analysis}
First, it is important to qualitatively address the behavior of each model. This will help us determine the sensitivity of each model in terms of atmospheric parameters inputs: temperature, pressure, and relative humidity. It is clear from Figure \ref{Fig:model_diff}, that all of the models behave in a similar way except for Zemax. In order to test the sensitivity of the models, we compare the results of a set of two models by varying the inputs as follow: i) temperature range from -5 to 25$^{\circ}$C; ii) pressure range from 700 to 800 mbar; and iii) relative humidity from 0 to 70 \%. We also assume the following default inputs: T = 13$^{\circ}$C, P = 740 mbar, RH = 20\%, and Z = 60$^{\circ}$. These ranges are the same ranges where the VLT telescopes operate. For demonstration purposes, we will show the comparison between the B \& P model as well as the Zemax model with respect to the Filippenko model (the other models return similar behavior as the ones presented in this subsection).

\subsubsection{B \& P vs Filippenko}
The left column of Figure \ref{Fig:models_variation}, represents the results of the comparison between the B \& P and Filippenko models regarding the temperature, pressure, and relative humidity variation. From these figures, we notice that the general difference between these models, due to the main refractivity formula (Equations \ref{eq:f_s}, \ref{eq:bp_s}), is negligible ($\sim$ 2.5 mas). We also notice that they have similar sensitivities in terms of temperature and pressure, while there is a variation of 1 mas (in the bluest part of the spectrum) due to the RH variation. 

\subsubsection{Zemax vs Filippenko}
\begin{figure*}
    \centering
    \includegraphics[width=\hsize]{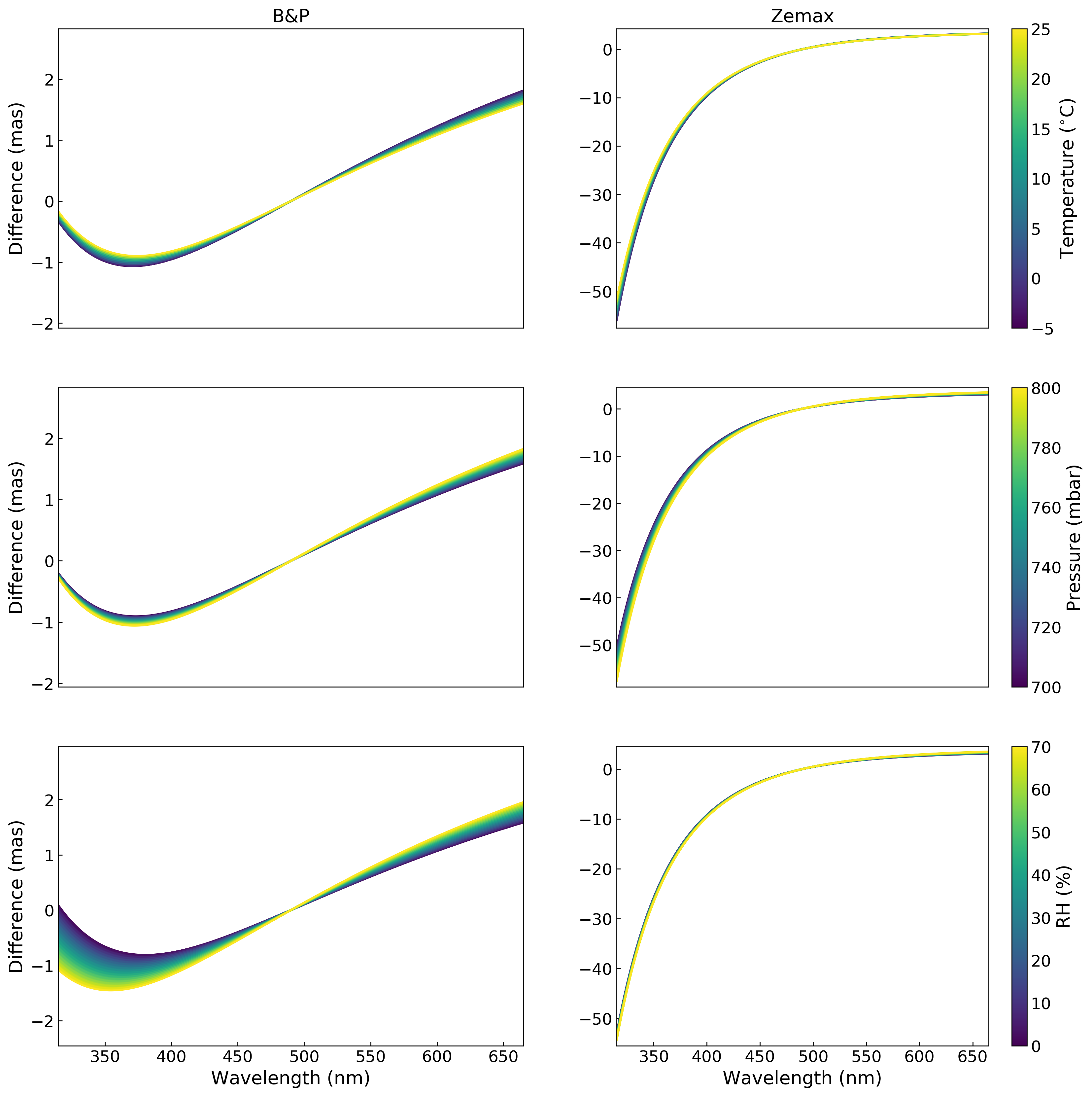}
    \caption{Left column: difference (in dispersion from the models) between the B \& P and Filippenko assuming the default inputs parameters: P = 740 mbar, RH = 20\%, and Z = 60$^{\circ}$. The temperature varies from -5 to 25 $^{\circ}$C as shown in the colorbar. The wavelength reference is assumed to be 490 nm; right: similar to left but comparing Zemax with Filippenko. We note that the scale on the y-axis between for B\&P and Zemax, is different.}
    \label{Fig:models_variation}
\end{figure*}

The right column of Figure \ref{Fig:models_variation}, represents the results of the comparison between the Zemax and Filippenko models regarding the temperature, pressure, and relative humidity. It is clear that the sensitivities of the models are similar, and the general difference is due to the main refractivity formula (Equations \ref{eq:f_s}, \ref{eq:z_s}).

\section{Method \& Observations}
\label{sec:method_observations}
\subsection{Method}
\label{subsec:method}
The method we developed to measure the atmospheric dispersion on-sky, is described in details in \citet{Wehbe2020b}. We limit ourselves here to mention the main steps only. We used the cross-dispersed slit spectrograph UVES \citep{Dekker2000}, with the slit directed along the dispersion direction and we removed the ADC from the optical path. This setup, allowed us to directly measure the atmospheric dispersion as a function of wavelength. The method is based on the determination of the position of the centroid of the spatial profile at each wavelength of each spectral order. Using the flat field frames as a reference point, we overlap the science frames and by applying cuts perpendicular to the spectral orders, we extract the centroids using a Gaussian and a Sigmoid function \citep[for more details see][]{Wehbe2020b}. In Figure 2 of \citet{Wehbe2020b}, we represent the overlapping of the two frames as well as the centroids extraction. By repeating the same procedure to all the spectral orders, through all the cuts, we get a plot that represents the variation of atmospheric dispersion in pixel as a function of wavelength. By using one of the two methods explained in \citet{Wehbe2020b}, we convert the atmospheric dispersion from pixels to mas in the sky. Using the method, we also were able to detect a non-negligible instrumental dispersion that should be corrected from the final results. The method shows an accuracy (see Table \ref{Table:uncertainty}), at the 1 $\sigma$ confidence level, of $\pm$ 18 mas in the blue and red ranges of UVES (315 to 665 nm).  

\begin{table}
    \caption{Uncertainties budget of the method, showing the main sources of errors in on-sky measurements \citep[adopted from Table 3 of][]{Wehbe2020b}.}
    \label{Table:uncertainty}
    \centering
    \begin{tabular}{c c}
        \hline
        Source of uncertainty & Contribution \\
        \hline
        Pixel scale & 0.87 mas \\
        Instrumental dispersion & 8.30 mas \\
        Atmospheric dispersion & 16.18 mas \\
         & \\
        \multicolumn{2}{c}{Accuracy of the method: 18 mas} \\
        \hline
    \end{tabular}
\end{table}

\subsection{Observations}
In order to perform an analysis over the optical range of the spectrum, we obtained observation from UVES, using a dichroic to cover the wavelength range of 303 nm to 665 nm. Our data was acquired based on observations made with UVES (program ID 4103.L-0942(A); PI: B. Wehbe). The spectrograph is equipped with two arms: i) blue in the 303-384 nm range, ii) red in the 487-567 nm and 590-665 nm (the red arm detector is a mosaic of two CCDs); the wavelength ranges are following our setup using a dichroic. In all observations, the slit of the spectrograph was oriented along the atmospheric dispersion direction, the ADC was off (outside of the optical path), and using dichroic 1 along with the two cross-dispersers (CD) 1 \& 3. In order to achieve the desired accuracy of $\sim$ 1 \% (on the dispersion measurements), 19 exposures of 20 seconds each were done for each target. We observed 3 targets between zenithal angles of 50$^{\circ}$ and 60$^{\circ}$, and one target between 14$^{\circ}$ and 19$^{\circ}$. Table \ref{Table:atm} summarizes the atmospheric conditions at the time of observation of each target. The targets chosen are O-type stars (HD17490 and HD150574), and B-type stars (HD143449 and HD165320). We choose O and B-type stars in order to achieve a higher signal-to-noise (S/N) ratios in particular in the blue range of the spectrum where the atmospheric dispersion is more severe. We also state that another important factor when choosing the targets was the zenithal angle at which it can be observed. The higher the angle, the bigger the atmospheric dispersion is. Two of these targets (HD117490 and HD143449) were observed with increasing zenithal angles between the first and last exposures. The other two targets (HD150574 and HD165320) were observed with decreasing zenithal angles between the first and last exposures. This is in particular importance because the atmospheric dispersion is in opposite directions between increasing and decreasing zenithal angles; an information important to take into consideration while correcting from the instrumental dispersion as it is considered as a fixed dispersion in quantity and direction.

\begin{table}
	\caption{Atmospheric conditions at the time of observation of each target.}             
	\label{Table:atm}      
	\centering
    \begin{tabular}{c c c c c}
        \hline
        Target & Zenithal angle & Temperature & Pressure & RH \\
         & $^{\circ}$ & $^{\circ}$C & mbar & \% \\
		\hline
        HD 117490 & 57.4 - 59.9 & 11.7 - 12.3 & 742.6 & 4.5 - 6 \\
        HD 150574 & 55.9 - 59.4 & 13.4 & 745 & 13.5 - 17 \\
        HD 143449 & 50 - 52.3 & 9.5 - 9.7 & 741.8 & 18.5 - 23 \\
        HD 165320 & 13.8 - 18.5 & 13.6 -13.8 & 744.7 & 14.5 - 16.5 \\
		\hline
	\end{tabular}	
\end{table}

\section{Data Analysis}
\label{sec:data_analysis}
As mentioned before, UVES is equipped with two arms: a blue one and a red one. Unfortunately, these two ranges do not overlap in our setup. This means that we can't evaluate the full range from 303 nm to 665 nm in one exposure. Therefore, we will divide our analysis in two subsections focusing on each arm separately. In order to reach the accuracy needed, we required 19 exposures of each target. It is important, when analyzing the results, to remove any outliers from the measurements. We first subtract the mean value from the residuals of each exposure. This will allow us to center the residuals around zero. This in fact is similar to using the guiding camera to center the target in the fiber. Then, we compute the half-peak-to-valley (HPTV) of the residuals of each exposure, for all the models of interest. Using the HPTV of all the exposures, we can identify the outliers using the Z-score method (Equation \ref{eq:Z_score}). Any exposure that return a Z-score higher than the threshold (Equation \ref{eq:threshold}), is considered as an outlier. Note that we repeat this procedure for each model. In the following part of the paper, we only take into account the exposures that return a Z-score below the threshold: all the exposures of HD117490 and HD165320, 11 exposures of HD150574, and 11 exposures of HD143449.  

\begin{equation}
    Z = \frac{x-\mu}{\sigma},
    \label{eq:Z_score}
\end{equation}
where x is the HPTV of each exposure, $\rm \mu$ is the PV average of all the exposures, and $\rm \sigma$ is the standard deviation.

\begin{equation}
    t = \frac{n-1}{\sqrt{n}},
    \label{eq:threshold}
\end{equation}
where n is the number of exposures of each target (19).

\subsection{The blue arm}
\label{subsec:blue}
\begin{figure}
	\centering
	\includegraphics[width=\hsize]{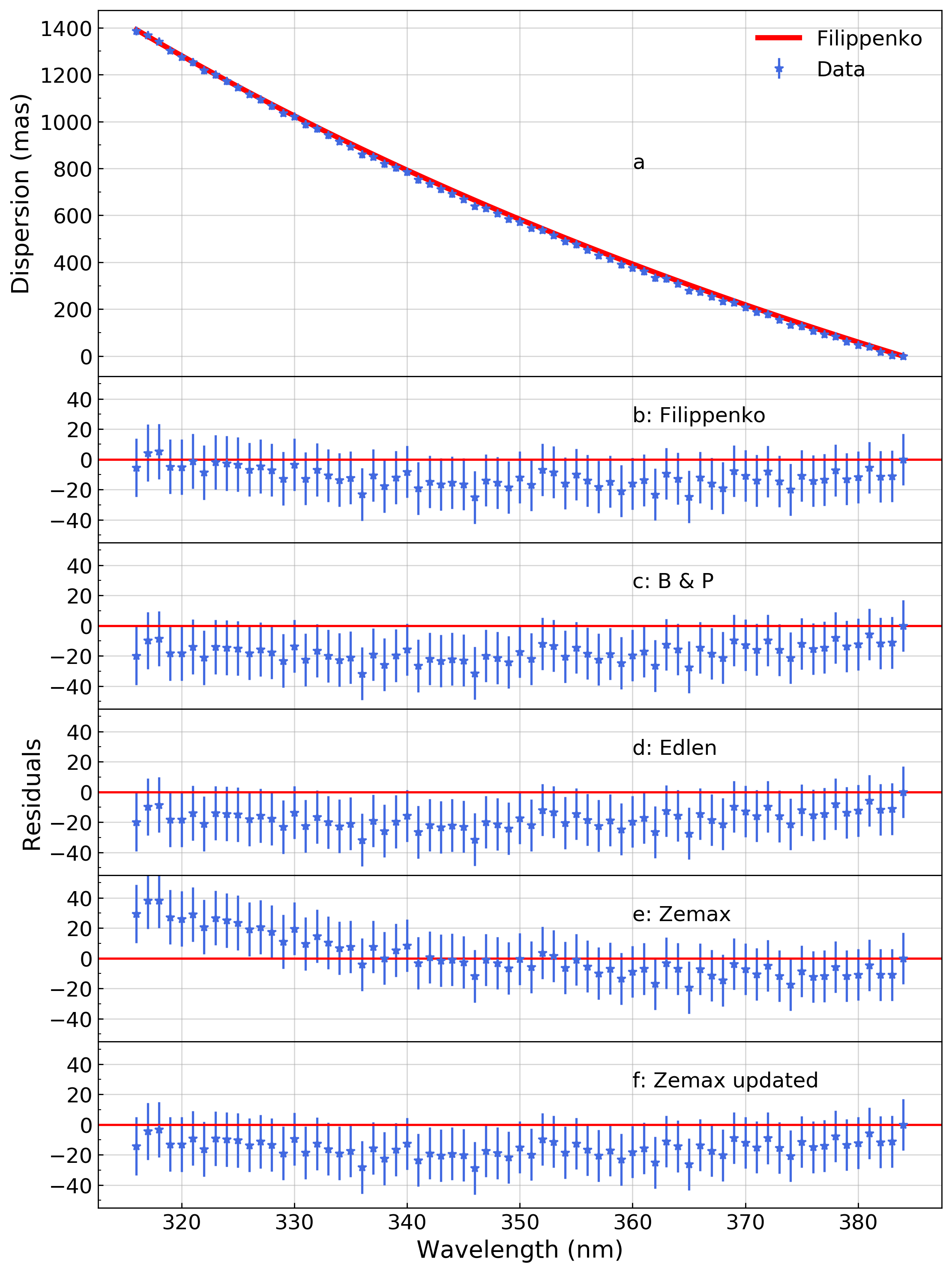}
	\caption{Panel a: measured atmospheric dispersion for one exposure of HD117490, overplotted by the expected dispersion from the Filippenko's model; panels b to f: atmospheric dispersion residuals for all the modesl of interest.}
	\label{Fig:blue_one}  
\end{figure}

\begin{figure*}
    \centering
    \includegraphics[width=\hsize]{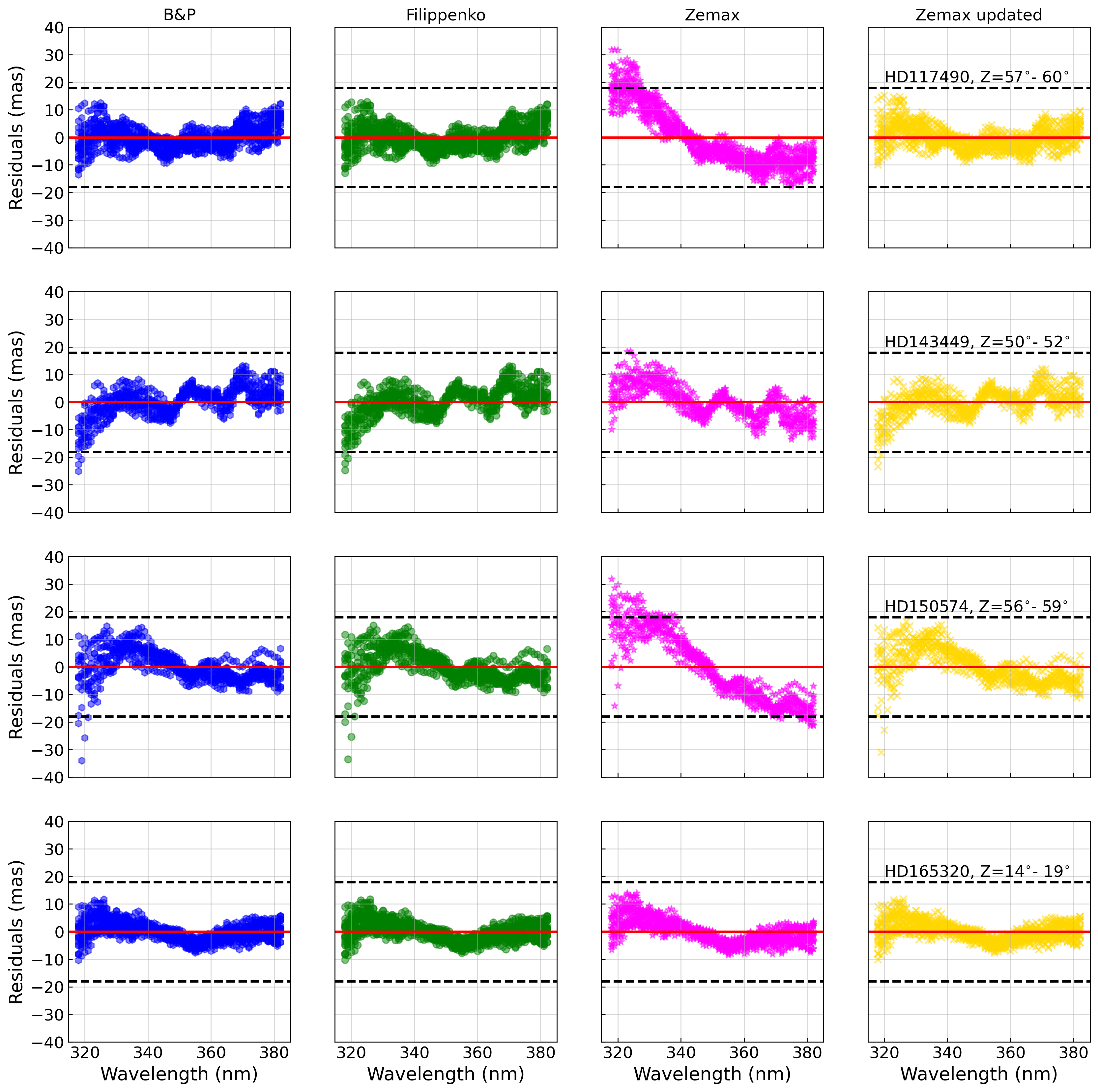}
    \caption{Residuals of all the exposures of all the targets for all the models of interest in the blue arm (315 - 385 nm). Each row represents a target, hence a zenithal angle. Each column represents a different atmospheric model. The two dashed lines represent the accuracy of the method: $\pm$ 18 mas.}
    \label{Fig:res_blue}
\end{figure*}

Similar to paper I, in our analysis we use the blue range starting 315 nm, since the points below are affected by low atmospheric transmission and low S/N ratios. In Figure \ref{Fig:blue_one}, we show the case of one exposure of the target HD117490. Panel \textbf{a} of Figure \ref{Fig:blue_one} shows the measured atmospheric dispersion as well as the computed one using the Filippenko's model. In the other panels (\textbf{b} to \textbf{f}) of Figure \ref{Fig:blue_one}, we show the residuals (i.e., data - model) for all the models of interest. The different models show similar behavior in terms of residuals, except for the Zemax model which seems to be underestimating the atmospheric dispersion in the bluest part of the range. To investigate more, we should look at the results of the 19 exposures of each target. After removing the outlier exposures, we plot the residuals of all the models. The Edlen model is excluded from these plots as it return similar results to the B \& P model. Figure \ref{Fig:res_blue} represents the results. It is clear from Figure \ref{Fig:res_blue}, that the models B \& P, Filippenko, and Zemax updated, are returning similar residuals for the 4 targets. It is also clear that they are able to return residuals between -10 and 10 mas, for all the zenithal angles tested (the 4 targets). Also from Figure \ref{Fig:res_blue}, we can see that the Zemax model is underestimating the atmospheric dispersion, especially in the bluest part of the spectrum. This is particularly clear in the highest zenithal angles cases (HD117490 and HD150574), where the dispersion is more severe. The updated version of the Zemax model introduced by \citet{Spano2014}, is taking these underestimation into consideration and returning better results in terms of residuals. The Zemax model should be avoided when dealing with the blue range, especially when expecting residuals below the 20 mas level. At lower zenithal angles, hence lower atmospheric dispersion (the case of HD165320 in Figure \ref{Fig:res_blue}), the 4 models return similar results between -10 and 10 mas.  

%
%


\subsection{The red arm}
\begin{figure}
	\centering
	\includegraphics[width=\hsize]{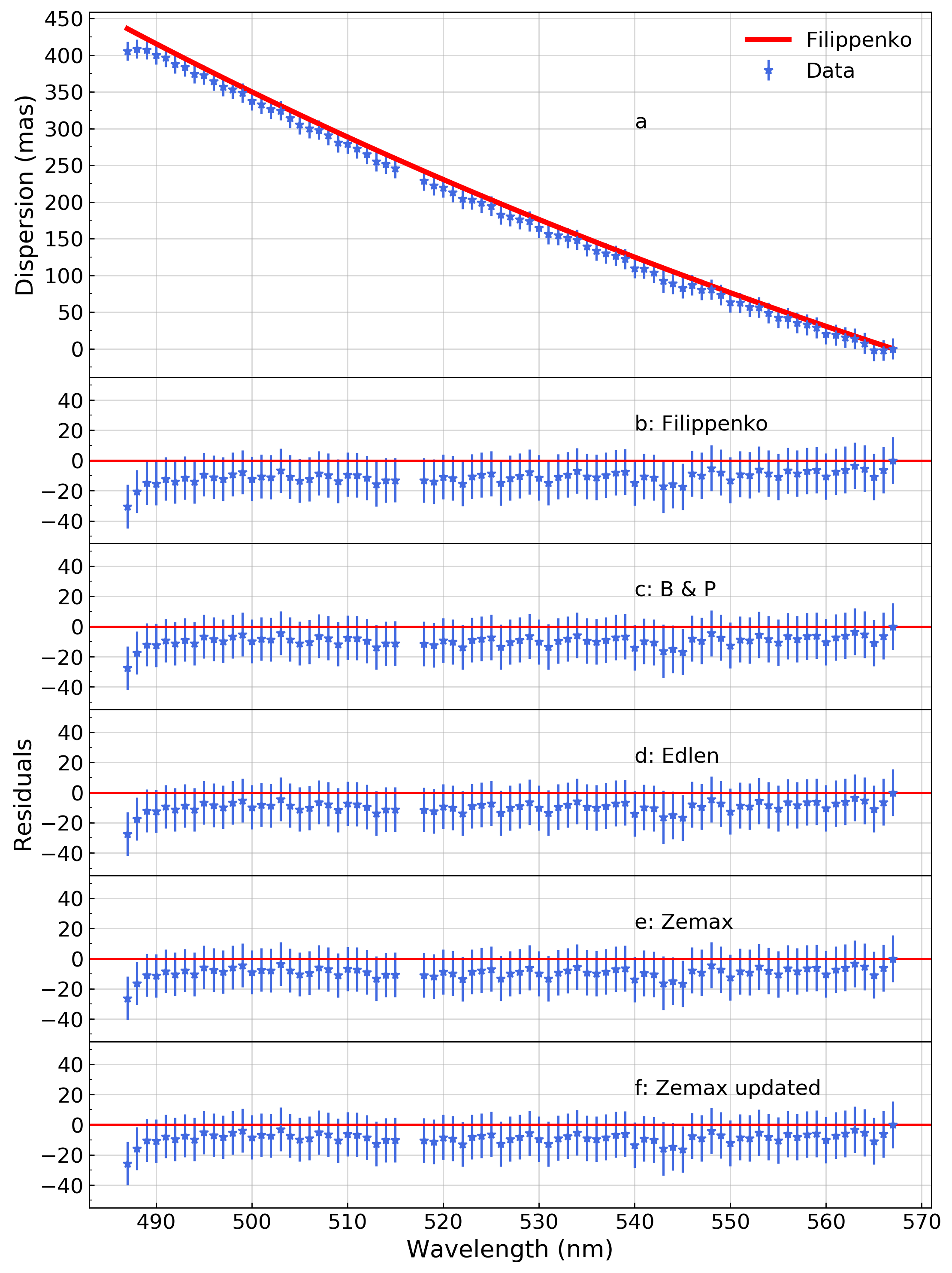}
	\caption{Similar to Figure \ref{Fig:blue_one}, for the wavelength range 487-567 nm.}
	\label{Fig:redl_one}  
\end{figure}

\begin{figure}
	\centering
	\includegraphics[width=\hsize]{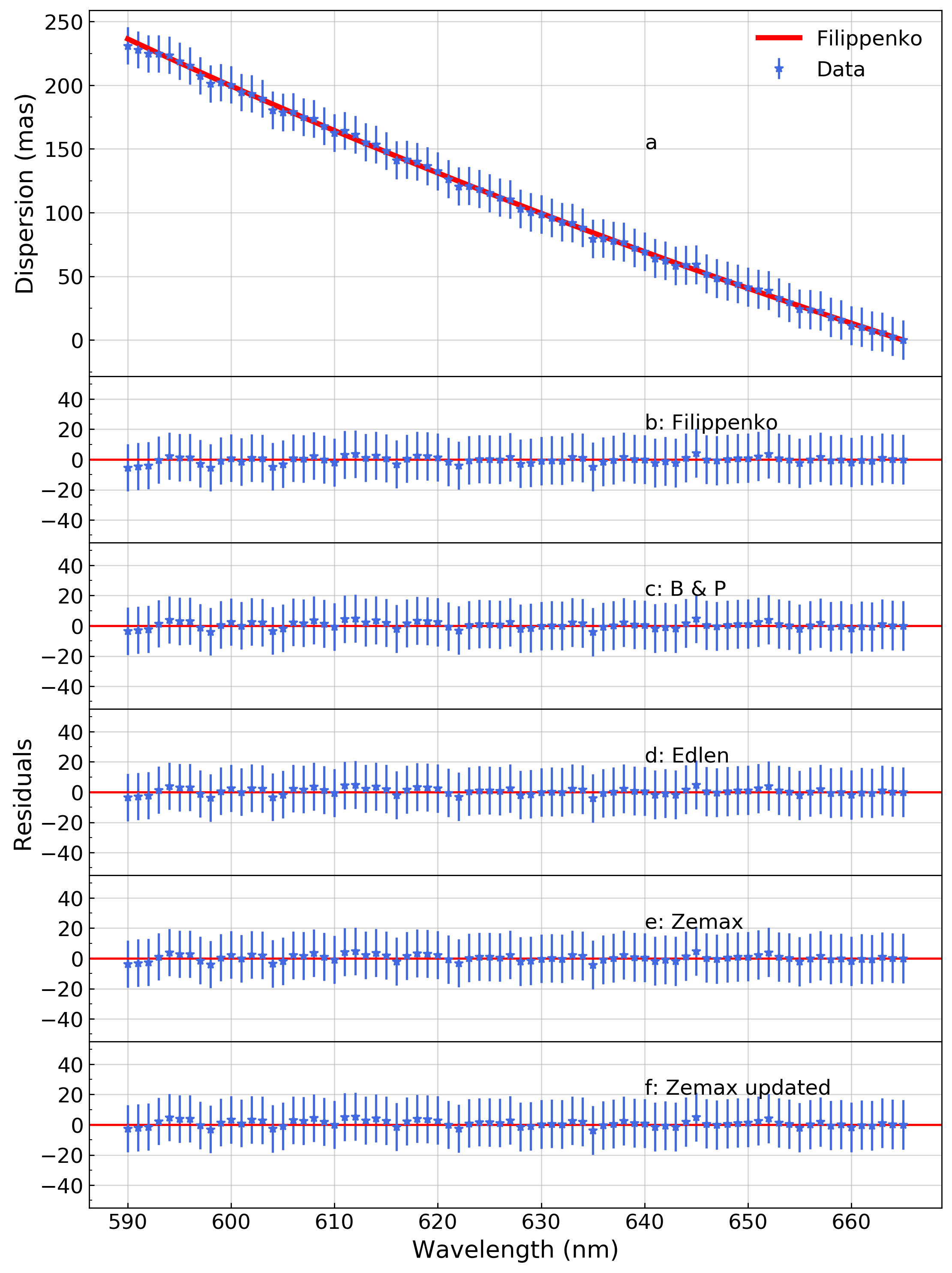}
	\caption{Similar to Figure \ref{Fig:blue_one}, for the wavelength range 590-665 nm.}
	\label{Fig:redu_one}  
\end{figure}

As it was mentioned before, the red arm detector is a mosaic of two CCDs resulting in two separate ranges: 487-567 nm, and $\rm 590-665 ~ nm$. In this subsection, we will show the results of the red arm in a similar way as presented in subsection \ref{subsec:blue}. In Figures \ref{Fig:redl_one}, and \ref{Fig:redu_one}, we show the case of one exposure of the target HD117490. Panel \textbf{a} of both figures shows the measured atmospheric dispersion as well as the computed one using the Filippenko's model, for the corresponding range. In the other panels (\textbf{b} to \textbf{f}), we show the residuals for all the models of interest. The different models show similar behavior in terms of residuals. In a similar manner to subsection \ref{subsec:blue}, we plot the residuals of all the exposures of all the targets. These residuals are shown in Figures \ref{Fig:res_redl} and \ref{Fig:res_redu} for the wavelength ranges of $\rm 487-567 ~ nm$ and 590-665 nm, respectively.

\subsubsection{The red range of 487-567 nm}
\begin{figure}
    \centering
    \includegraphics[width=\hsize]{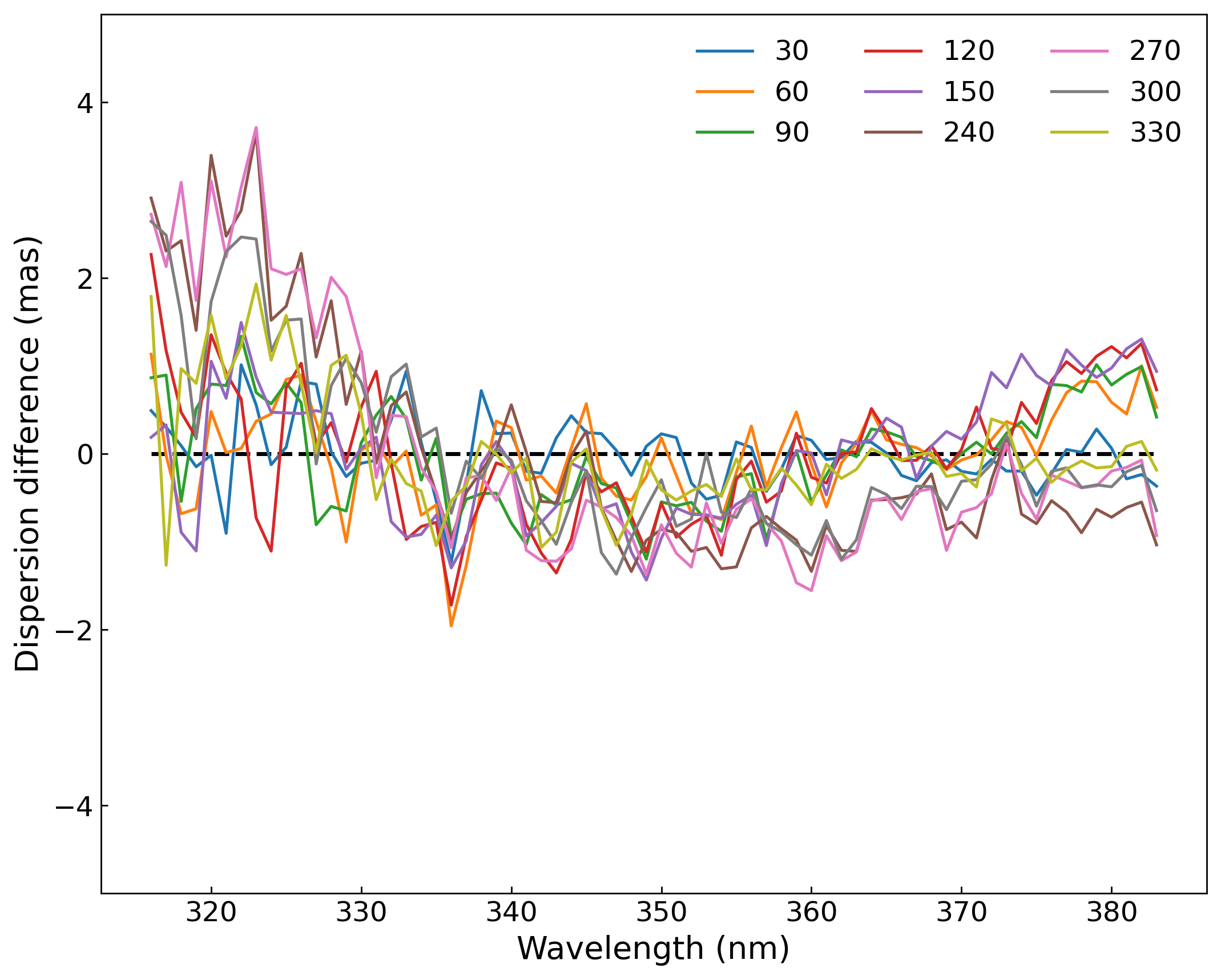}
    \includegraphics[width=\hsize]{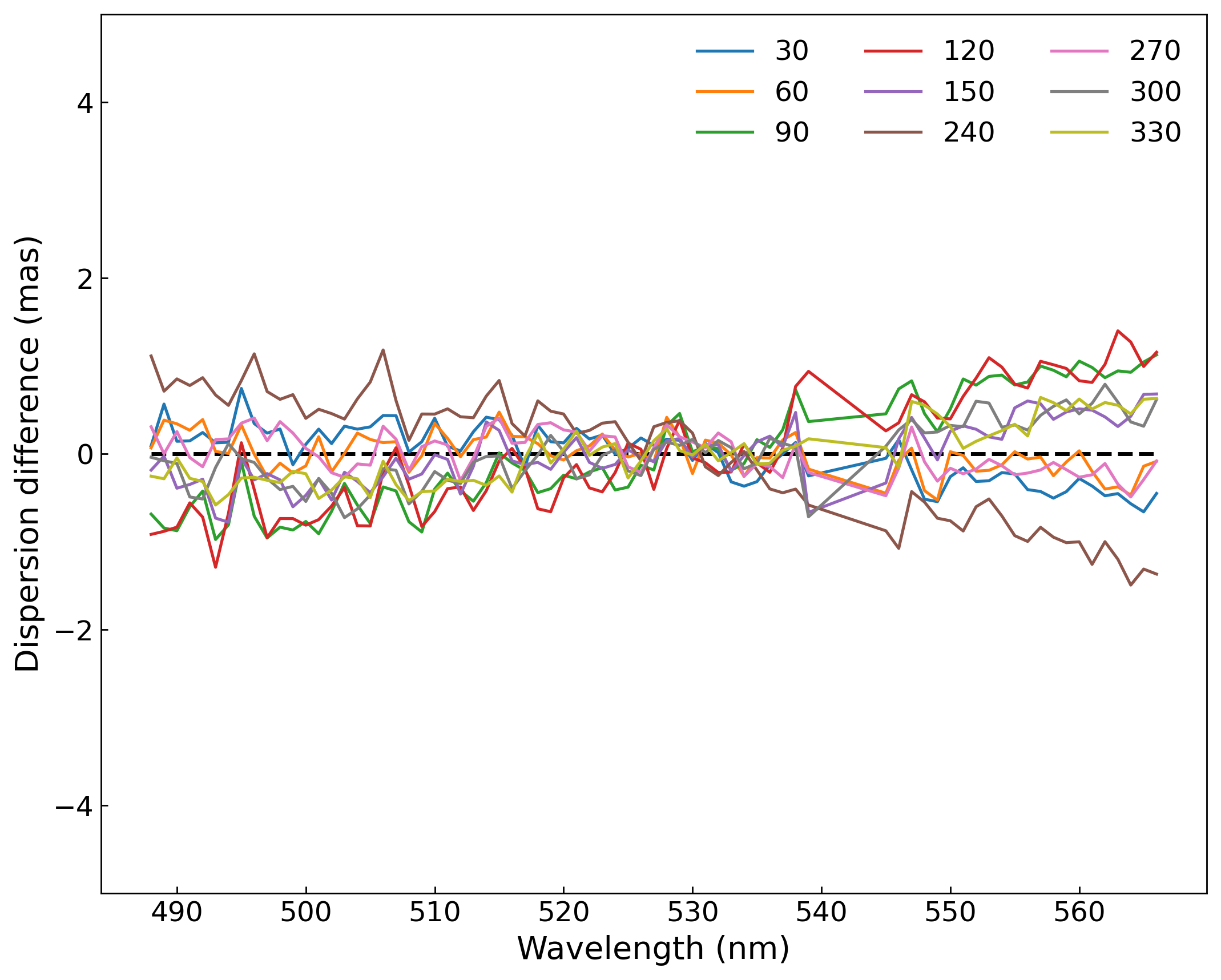}
    \includegraphics[width=\hsize]{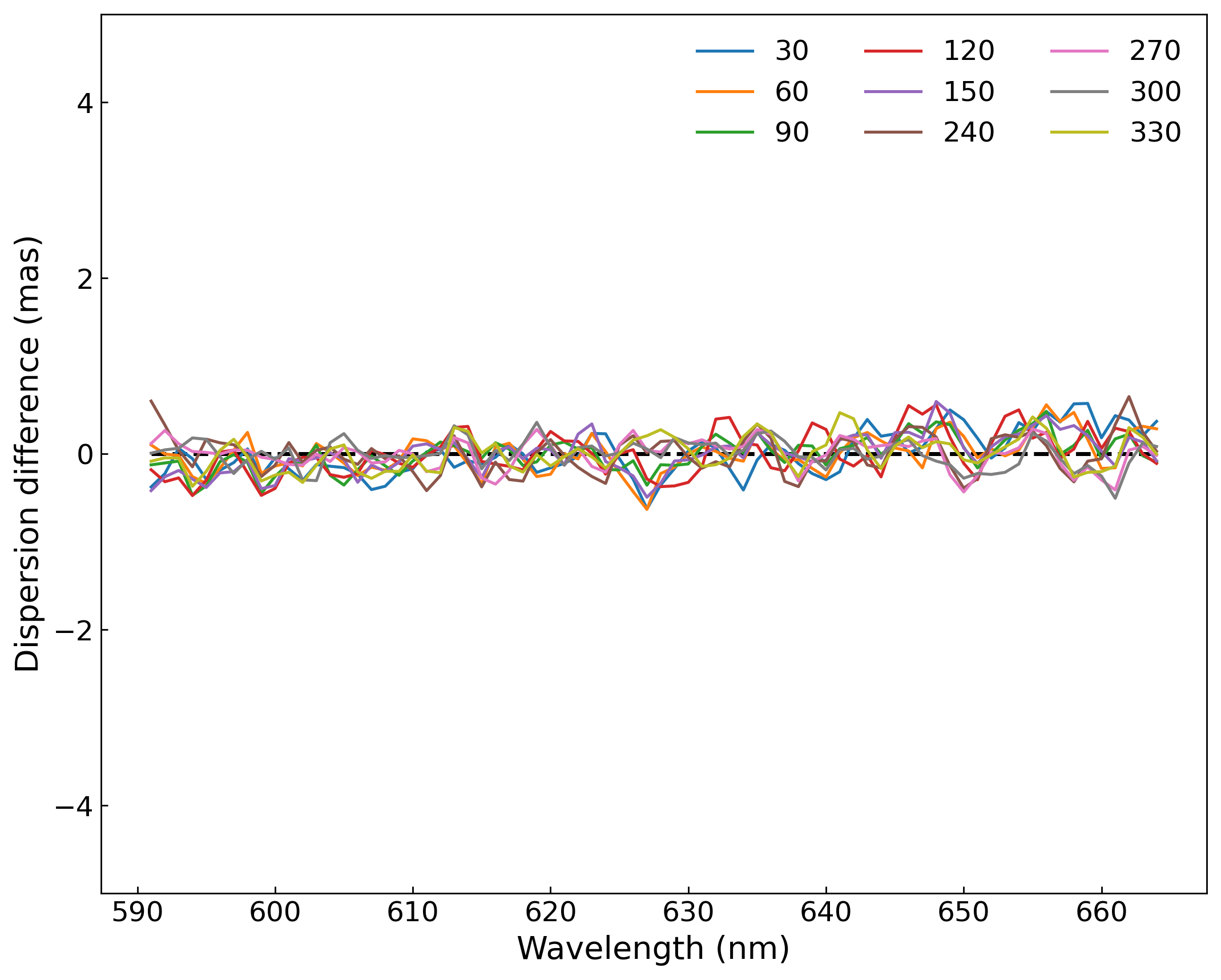}
    \caption{Effect of derotator angle on the measured atmospheric dispersion. The frame at derotator position 0$^{\circ}$ is used as a reference frame for the comparison. Top: for the blue range of 315-384 nm, middle: for the red range of 487-567 nm, bottom: for the red range of 590-665 nm.}
    \label{Fig:drotvar}
\end{figure}

\begin{figure*}
    \centering
    \includegraphics[width=\hsize]{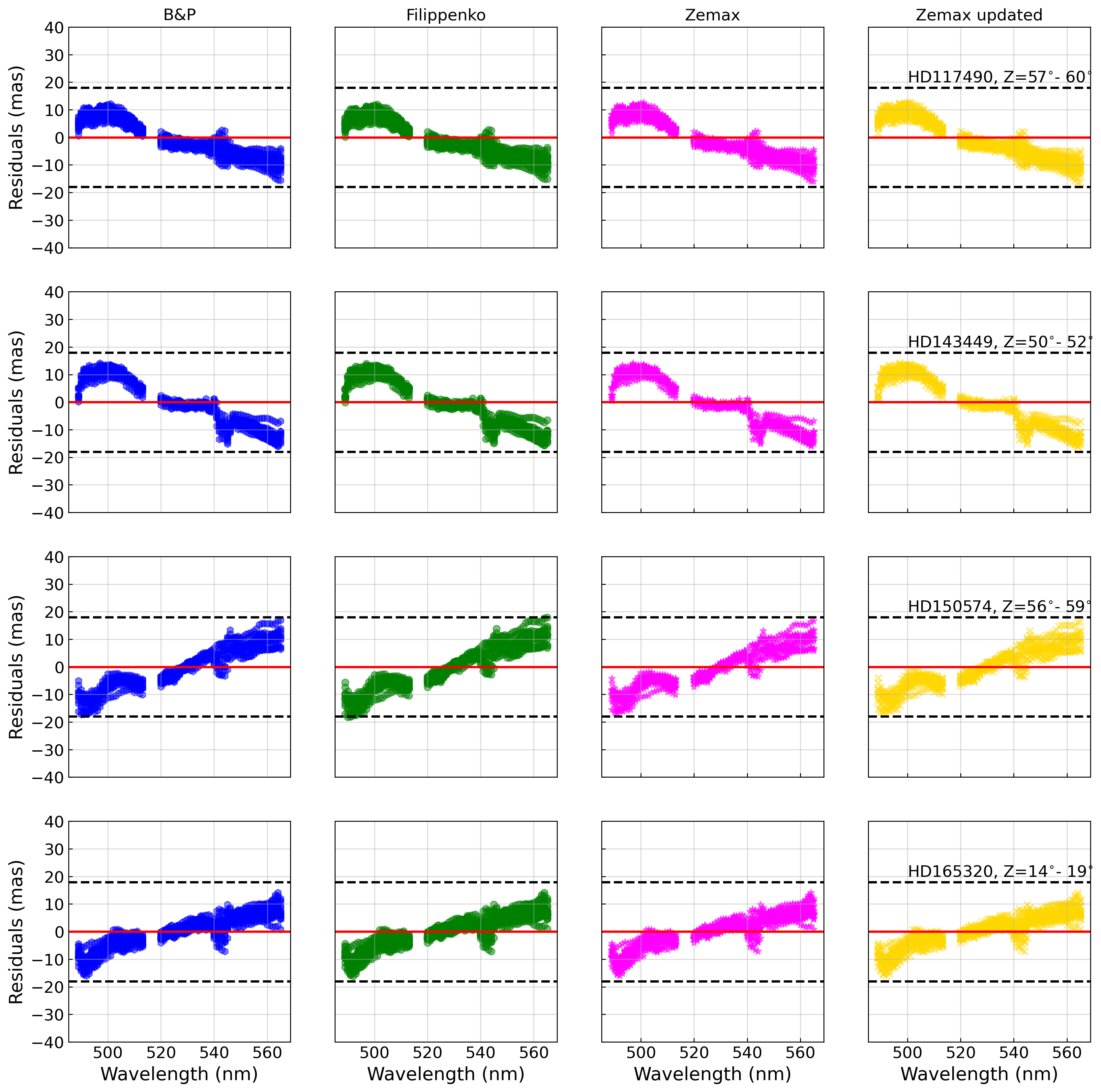}
    \caption{Similar to Figure \ref{Fig:res_blue}, but for the wavelength range of 487-567 nm.}
    \label{Fig:res_redl}
\end{figure*}

The residuals of all the models of all the targets are within the uncertainty level as seen in Figure \ref{Fig:res_redl}. In a first analysis, we can consider them negligible. Nevertheless, looking at Figure \ref{Fig:res_redl} and not taking into consideration the uncertainty level, a slope in the residuals is clear. As this effect is withing our uncertainty level, it does not affect our results. However, we tried to investigate the possible cause. \\
This effect is model independent which rule out the idea of an error in the models. It is also target independent. The four targets were observed at different days, different zenithal angles, but with the same optical setup of the spectrograph. This indicates that it is an instrumental effect, and not target related, that we were not able to detect using the airmass 1 data.  \\
One possible cause to explain this slope could be the derotator, affecting our calibration between the science and flat frames used. To investigate this effect, we requested a set of flat frames\footnote{This test was executed as an instrumental test and as such it’s not available from the ESO archive. The data is available upon request from the author.} while rotating the derotator of UVES between 0$^{\circ}$ and 360$^{\circ}$, with steps of 30$^{\circ}$. The angles of 180$^{\circ}$ and 210$^{\circ}$ were excluded from the set due to technical issues. By measuring the atmospheric dispersion using different flat frames with different derotator positions, we were able to quantify the effect of the derotator on the instrumental dispersion. We use the frame of 0$^{\circ}$ as a reference point. Figure \ref{Fig:drotvar} shows the effect of the derotator position on the measured atmospheric dispersion.

The effect of the derotator is at the level of $\pm$ 2 mas in the blue range, and $\pm$ 1 mas in the red range (as seen in Figure \ref{Fig:drotvar}) which rule out the responsibility of the derotator. Another possible cause could be a small miss-alignment variation in the optics. It is important to state here that this effect is so small that it will not affect the science observations of UVES. However, since we are trying to measure residuals at the level of few mas, we are able to detect it. The set of the flat frames was taken in two nights (27 and 28 of December 2020) where there was an earthquake near Paranal where UVES is installed. By comparing the zero flat frame before and after the earthquake, we can quantify a linear shift in the dispersion at the level of 10 mas in the blue range and 3 mas in the red range. This shift, even though at a different level to what we are seeing in Figure \ref{Fig:res_redl}, shows a similar behavior. This might explain the residuals linear behavior. In fact, the airmass 1 set of data used to quantify the instrumental dispersion was taken in 2015, while our observations were done in 2019. Any small misalignment variation on the red mosaic chip, or even an earthquake, might introduce such an effect. Another possible cause could be related to thermal variations from observation to observation that might create some offsets on the data. Unfortunately, we were not able to find a single cause that can confirm the presence of this slope. We suspect that it could be due to a sum of several small effects, that don't affect the UVES science observations, but clear in our analysis. \\
Even though there is an extra dispersion that it is unclear for us from where, we are confident from our results as this slope effect is within our uncertainty level. It has no impact on our conclusions. The residuals of all the targets are between the -10 and 10 mas, similar to the blue range.    

\subsubsection{The red range of 590-665 nm}
\begin{figure*}
    \centering
    \includegraphics[width=\hsize]{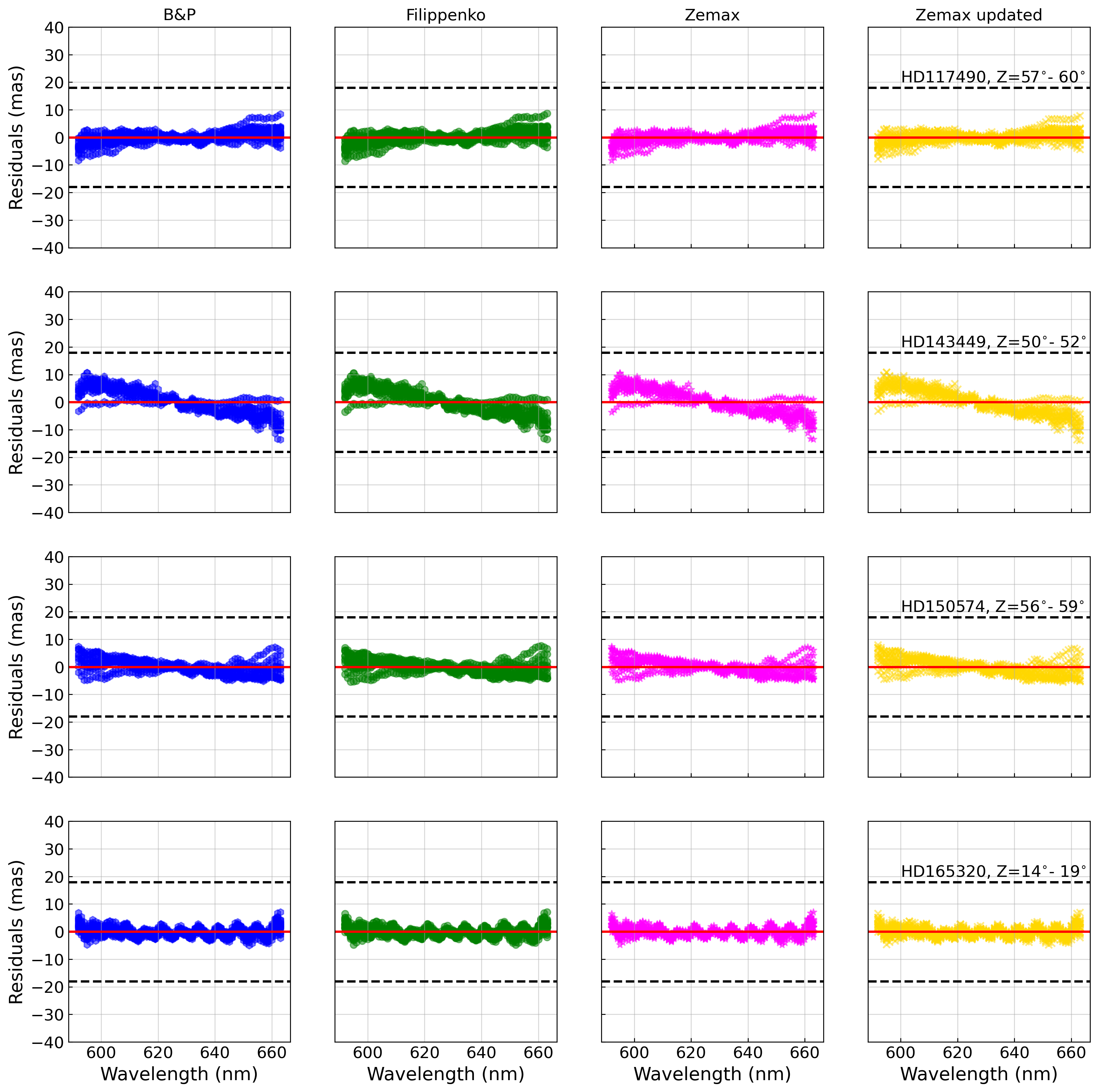}
    \caption{Similar to Figure \ref{Fig:res_blue}, but for the wavelength range of 590-665 nm.}
    \label{Fig:res_redu}
\end{figure*}

The residuals of the range 590-665 nm are between -5 and 5 mas for all the targets and all the models, as seen in Figure \ref{Fig:res_redu}. The residuals are smaller in this range from the fact that the atmospheric dispersion is smaller and the difference between the models at this level is almost negligible. All the models are able to return residuals this low, independent of the zenithal angles. The slope detected in HD143449 in Figure \ref{Fig:res_redu}, is due to errors in our measurements. They are at the level few mas and can be considered negligible. The periodicity seen in the case of HD165320 is due to the order separation. It is particularly seen in this range since the level of atmospheric dispersion, hence the residuals, is really small compared to the other red range and to the blue range.

%
%
%
%
%

\section{Conclusions}
\label{sec:conclusions}
We present an analysis of different atmospheric dispersion models used during the design phase of an ADC. Using the method developed by \citet{Wehbe2020b}, we were able to compare the models of interest to on-sky measurements. To the best of our knowledge, this is the first time that atmospheric models are being compared to on-sky measurements, in particular in the blue part of the spectra, where the atmospheric dispersion is more severe. Even though the method we used has an accuracy of 18 mas, we were able to compare the different models. We can notice that in the blue part of the range, the Zemax model is returning higher residuals than the other models. While in the red ranges, the results are similar. From this paper, we find the following:

\begin{enumerate}
    \item The validation of the method is successfully extended from paper I to four different targets covering different ranges of atmospheric parameters and in particular different zenithal angles.
    \item It is clear that the Zemax updated model is better than the Zemax model, in terms of estimating the on-sky dispersion. We conclude this from the fact that the residuals are lower, in particular in the blue range of the spectra where it is well known by now that the Zemax model is underestimating the atmospheric dispersion.
    \item In the blue range of the spectra, and when expecting residuals at the level of few tens of mas, we recommend to avoid using the Zemax model. 
    \item In the red range of the spectra, all the models we tested return similar results. Any of the models can be safely used when expecting residuals at the level of few tens of mas.
    \item The effect seen in Figure \ref{Fig:res_redl}, is to our estimation, due to an extra instrumental dispersion not detected in the data of airmass 1. Even that the amplitude is below our method accuracy, the effect is clear. Ruling out the effect of the derotator position, strengthen the accuracy of our method. We suspect that it could be to a miss-alignment in the red mosaic chip. When dealing with the range of 487-567 nm, we reached the method's limits ($\pm$ 18 mas) . Even by using UVES, one of the state-of-the-art spectrographs, it is not easy to quantify the atmospheric dispersion below the accuracy of our method. This only shows that measuring on-sky the atmospheric dispersion is a challenging task. 
\end{enumerate}

%
%

\section*{Acknowledgements}
The first author is supported by a Funda\c{c}\~{a}o para a Ci\^{e}ncia e Tecnologia (FCT) fellowship (PD/BD/135225/2017), under the FCT PD Program PhD::SPACE (PD/00040/2012). This work was supported by FCT/MCTES through national funds and by FEDER - Fundo Europeu de Desenvolvimento Regional through COMPETE2020 - Programa Operacional Competitividade e Internacionaliza\c{c}\~{a}o by these grants: UID/FIS/04434/2019; PTDC/FIS-AST/32113/2017 \& POCI-01-0145-FEDER-032113.

\section*{Data availability}
The data underlying this article are based on observations made with ESO Telescopes at the Paranal Observatory under program ID 4103.L-0942(A) (with the UVES spectrograph at the ESO VLT UT2 telescope) and available in ESO Science Archive Facility at http://archive.eso.org/, and can be accessed with the program ID. The flat frames used in the derotator test are available upon request from the author.




\bibliographystyle{mnras}
\bibliography{wehbereferences} 

\begin{thebibliography}{}
\makeatletter
\relax
\def\mn@urlcharsother{\let\do\@makeother \do\$\do\&\do\#\do\^\do\_\do\%\do\~}
\def\mn@doi{\begingroup\mn@urlcharsother \@ifnextchar [ {\mn@doi@}
  {\mn@doi@[]}}
\def\mn@doi@[#1]#2{\def\@tempa{#1}\ifx\@tempa\@empty \href
  {http://dx.doi.org/#2} {doi:#2}\else \href {http://dx.doi.org/#2} {#1}\fi
  \endgroup}
\def\mn@eprint#1#2{\mn@eprint@#1:#2::\@nil}
\def\mn@eprint@arXiv#1{\href {http://arxiv.org/abs/#1} {{\tt arXiv:#1}}}
\def\mn@eprint@dblp#1{\href {http://dblp.uni-trier.de/rec/bibtex/#1.xml}
  {dblp:#1}}
\def\mn@eprint@#1:#2:#3:#4\@nil{\def\@tempa {#1}\def\@tempb {#2}\def\@tempc
  {#3}\ifx \@tempc \@empty \let \@tempc \@tempb \let \@tempb \@tempa \fi \ifx
  \@tempb \@empty \def\@tempb {arXiv}\fi \@ifundefined
  {mn@eprint@\@tempb}{\@tempb:\@tempc}{\expandafter \expandafter \csname
  mn@eprint@\@tempb\endcsname \expandafter{\@tempc}}}

\bibitem[\protect\citeauthoryear{{Barrell} \& {Sears}}{{Barrell} \&
  {Sears}}{1939}]{Barrell1939}
{Barrell} H.,  {Sears} J.,  1939, \mn@doi [Philosophical Transactions of the
  Royal Society of London Series A] {10.1098/rsta.1939.0004}, \href
  {https://ui.adsabs.harvard.edu/abs/1939RSPTA.238....1B} {238, 1}

\bibitem[\protect\citeauthoryear{{Bechter}, {Bechter}, {Crepp}, {King}  \&
  {Crass}}{{Bechter} et~al.}{2018}]{Bechter2018}
{Bechter} A.~J.,  {Bechter} E.~B.,  {Crepp} J.~R.,  {King} D.,   {Crass} J.,
  2018, in \procspie. p. 107026T (\mn@eprint {arXiv} {1812.02704}),
  \mn@doi{10.1117/12.2313658}

\bibitem[\protect\citeauthoryear{{Blackman}, {Ong}  \& {Fischer}}{{Blackman}
  et~al.}{2019}]{Blackman2019}
{Blackman} R.~T.,  {Ong} J.~M.~J.,   {Fischer} D.~A.,  2019, \mn@doi [\aj]
  {10.3847/1538-3881/ab24c3}, \href
  {https://ui.adsabs.harvard.edu/abs/2019AJ....158...40B} {158, 40}

\bibitem[\protect\citeauthoryear{{B{\"o}nsch} \& {Potulski}}{{B{\"o}nsch} \&
  {Potulski}}{1998}]{Bonsch1998}
{B{\"o}nsch} G.,  {Potulski} E.,  1998, \mn@doi [Metrologia]
  {10.1088/0026-1394/35/2/8}, \href
  {http://adsabs.harvard.edu/abs/1998Metro..35..133B} {35, 133}

\bibitem[\protect\citeauthoryear{{Cabral} et~al.,}{{Cabral}
  et~al.}{2012}]{Cabral2012}
{Cabral} A.,  et~al., 2012, in Ground-based and Airborne Telescopes IV. p.
  84444F, \mn@doi{10.1117/12.926093}

\bibitem[\protect\citeauthoryear{{Dekker}, {D'Odorico}, {Kaufer}, {Delabre}  \&
  {Kotzlowski}}{{Dekker} et~al.}{2000}]{Dekker2000}
{Dekker} H.,  {D'Odorico} S.,  {Kaufer} A.,  {Delabre} B.,   {Kotzlowski} H.,
  2000, {Design, construction, and performance of UVES, the echelle
  spectrograph for the UT2 Kueyen Telescope at the ESO Paranal Observatory}.
pp 534--545, \mn@doi{10.1117/12.395512}

\bibitem[\protect\citeauthoryear{{Edlen}}{{Edlen}}{1953}]{Edlen1953}
{Edlen} B.,  1953, Journal of the Optical Society of America (1917-1983), \href
  {https://ui.adsabs.harvard.edu/abs/1953JOSA...43..339E} {43, 339}

\bibitem[\protect\citeauthoryear{{Edl{\'e}n}}{{Edl{\'e}n}}{1966}]{Edlen1966}
{Edl{\'e}n} B.,  1966, \mn@doi [Metrologia] {10.1088/0026-1394/2/2/002}, \href
  {http://adsabs.harvard.edu/abs/1966Metro...2...71E} {2, 71}

\bibitem[\protect\citeauthoryear{{Filippenko}}{{Filippenko}}{1982}]{Filippenko1982}
{Filippenko} A.~V.,  1982, \mn@doi [\pasp] {10.1086/131052}, \href
  {http://adsabs.harvard.edu/abs/1982PASP...94..715F} {94, 715}

\bibitem[\protect\citeauthoryear{{Hohenkerk} \& {Sinclair}}{{Hohenkerk} \&
  {Sinclair}}{1985}]{Zemax}
{Hohenkerk} C.~Y.,  {Sinclair} A.~T.,  1985, NAO Tech. Note No. 63

\bibitem[\protect\citeauthoryear{{Skemer} et~al.,}{{Skemer}
  et~al.}{2009}]{Skemer2009}
{Skemer} A.~J.,  et~al., 2009, \mn@doi [\pasp] {10.1086/605312}, \href
  {http://adsabs.harvard.edu/abs/2009PASP..121..897S} {121, 897}

\bibitem[\protect\citeauthoryear{{Span{\`o}}}{{Span{\`o}}}{2014}]{Spano2014}
{Span{\`o}} P.,  2014, in Advances in Optical and Mechanical Technologies for
  Telescopes and Instrumentation. p. 915157, \mn@doi{10.1117/12.2057072}

\bibitem[\protect\citeauthoryear{{Wehbe}, {Cabral}, {Martins}, {Figueira},
  {Santos}  \& {{\'A}vila}}{{Wehbe} et~al.}{2020a}]{Wehbe2020a}
{Wehbe} B.,  {Cabral} A.,  {Martins} J.~H.~C.,  {Figueira} P.,  {Santos} N.~C.,
    {{\'A}vila} G.,  2020a, \mn@doi [\mnras] {10.1093/mnras/stz3256}, \href
  {https://ui.adsabs.harvard.edu/abs/2020MNRAS.491.3515W} {491, 3515}

\bibitem[\protect\citeauthoryear{{Wehbe}, {Cabral}  \& {{\'A}vila}}{{Wehbe}
  et~al.}{2020b}]{Wehbe2020b}
{Wehbe} B.,  {Cabral} A.,   {{\'A}vila} G.,  2020b, \mn@doi [\mnras]
  {10.1093/mnras/staa2726}, \href
  {https://ui.adsabs.harvard.edu/abs/2020MNRAS.499..183W} {499, 183}

\makeatother
\end{thebibliography}








\bsp	
\label{lastpage}
\end{document}